\documentclass[onecolumn]{article}

\usepackage{graphicx}
\usepackage{ amssymb}
\usepackage[ansinew]{inputenc}
\usepackage{authblk}

\begin{document}

\title{A model for the interaction of high-energy particles in straight and bent 
crystals implemented in Geant4}

\author[1]{E. Bagli} 
\author[2]{M. Asai}
\author[3]{D. Brandt}
\author[2]{A. Dotti}
\author[1]{V. Guidi}
\author[2]{D.H. Wright}

\affil[1]{INFN Sezione di Ferrara, Dipartimento di Fisica e Scienze della Terra, Universit{\`a} di Ferrara Via Saragat 1, 44122 Ferrara, Italy}
\affil[2]{SLAC National Accelerator Laboratory, 2575 Sand Hill Rd, Menlo Park, CA 94025, United States of America}
\affil[3]{ Fasanenstr. 126, 82008 Unterhaching, Germany}

\maketitle

\begin{abstract}
A model for the simulation of orientational effects in straight and bent periodic
atomic structures is presented. The continuum potential approximation has been adopted.
The model allows the manipulation of particle 
trajectories by means of straight and bent crystals and the scaling of the cross 
sections of hadronic and electromagnetic processes for channeled particles. Based
on such a model, an extension of the Geant4 toolkit has been developed.  The code 
has been validated against data from channeling experiments carried out at CERN.
\end{abstract}

\maketitle
\newpage
\section*{Introduction}
The interaction of either charged or neutral particles with crystals is an area of
science under development.  Coherent effects of ultra-relativistic particles in 
crystals allow the manipulation of particle trajectories thanks to the strong 
electric field generated between crystal planes and axes 
\cite{Tsyganov682,Tsyganov684,Taratin1987425}.  Important examples of the 
interaction of neutral particles in crystals include production of 
electron-positron pairs and birefringence of high energy gamma 
quanta \cite{Okazaki2000110,1999hep.ex....4029M,PhysRevD.88.053009}.
Radiation emission due to curved trajectories in bent crystals has been seen to 
enhance photon production through bremsstrahlung, channeling radiation, 
parametric X-ray radiation, undulators
\cite{TerMikaelian,LandauLifshitz,Akhiezer:Shulga,BaierKatkov,doi:10.1142/S0218301304002557} 
and recently through volume reflection and multiple volume reflection 
\cite{1748-0221-3-02-P02005,PhysRevA.86.042903}. 
The inelastic nuclear interaction rate is known to be modified by channeling and
volume reflection \cite{Scandale20102655}.

Various applications of orientational phenomena with crystals have been proposed 
and investigated such as
\begin{itemize}
\item beam steering, \cite{Elishev1979387}
\item extraction and collimation in circular accelerators and \cite{PhysRevLett.87.094802,PhysRevSTAB.5.043501,FlillerIII200547,Scandale201078}
\item splitting and focusing of external beams \cite{Denisov1992382}.
\end{itemize}

Bent crystals have also been proposed as beam collimators
\cite{Scandale:1357606} and extractors
\cite{Fidecaro:220522,Jensen:291236,Rakotozafindrabe:1492806,Lansberg:1503414} for the LHC.
Indeed, with recent optimizations in manufacturing techniques \cite{0022-3727-41-24-245501,baricordi:061908}
and crystal holders \cite{ipac10tors}, bent crystals have been produced with record deflection 
efficiencies \cite{Scandale2009129}.
As a consequence of this and the reduction of the nuclear interaction rate for channeled positive particles
\cite{Scandale20102655}, the use of collimation systems based on bent crystals has proven to lower beam 
losses throughout
the SPS synchrotron for protons \cite{Scandale2012231} and for Pb ions \cite{PhysRevLett.79.4182,Scandale2011547}. 
 
The study of coherent effects for the interaction of particles with aligned structures have
always exploited opportunities furnished by numerical simulations with the most advanced computers and 
computational methods of the current period.  
Various approaches have been adopted.

The binary collision model allows the determination of
the trajectory of a low energy particle in a crystal with high precision, but it
is computationally expensive due to the need to solve the equation of motion of a
particle with an integration step smaller than the cell distance between two 
neighboring atoms, which is typically less than $1 {\AA}$. As an example,
the Monte Carlo code by Oen and Robinson \cite{PhysRev.132.2385}  was
capable of predicting the experimental results observed in 1963
\cite{PhysRevLett.10.399}.

By adopting the continuum 
approximation \cite{Dansk.Fys.34.14}, the equation of motion can be solved in one dimension for planar
channeling with an integration step of up to 1 ${\mu}m$ for GeV particles
\cite{FLUX7,Xavier1990278,taratin98,PhysRevE.51.3522,refId0,Bagli2013124}, with a
high computational cost for each particle due to the necessity of integrating over
the full particle trajectory. As an example of the capability of such a method,
in 1987 Vorobiev and Taratin predicted the volume reflection phenomenon in bent 
crystals \cite{Taratin1987425} which was first observed in 2006 by the H8RD22 
collaboration \cite{PhysRevLett.97.144801}. In 2013, an approach based on the numerical 
integration of classical relativistic equations of motion in a  dynamical generation 
was developed for the MBN Explorer software package in order to study relativistic 
phenomena in various environments such as crystals, amorphous bodies and  
biological media \cite{Sushko2013404}. 
A Fluka model for the simulation of planar channeling of positive particles in
bent crystals relies on the continuum potential approximation was proposed in 2013 \cite{Schoofs2013115}.

Thanks to the large amount of data \cite{PhysRevLett.98.154801,PhysRevLett.101.234801,Scandale2008109,PhysRevLett.102.084801,PhysRevSTAB.11.063501,PhysRevA.79.012903,Scandale2009129,Scandale2009233,Scandale2009301,Scandale2010284,Scandale2010545,Scandale20102655,Scandale2011180,salvador:234102,1748-0221-7-04-P04002,Scandale201370,PhysRevLett.110.175502}
with track reconstruction resolutions of $<10{\mu}$rad,
\cite{PhysRevLett.101.234801} Monte Carlo codes based
on the experimental cross sections of orientational phenomena were developed
\cite{Hasan2010449,1748-0221-7-04-P04002}.  With this model, very high 
computational throughput is achieved but the scaling of dechanneling models
 is inaccurate due to the lack of a dedicated campaign of measurement. 
 Moreover such an approach is not  suitable to describe the cross section
  variation of physical phenomena for channeled particles.

Nowadays Monte Carlo simulations of the interaction of particles with matter are 
usually done with download-able toolkits such as Geant4 \cite{Agostinelli2003250}
and Fluka \cite{Ferrari:898301}. Such Monte Carlo codes are continuously expanded
and improved thanks to the collaborative effort of scientists from around the 
world.  Geant4, an object-oriented toolkit, has seen a large expansion of its user
community in recent years.  As an example, applications simulated by Geant4 range
from particle transportation in the ATLAS detector 
\cite{doi:10.114297898127736780090} to calculations of dose distribution curves
for a typical proton therapy beam line \cite{5402279}, and from radiation analysis
for space instruments \cite{1589198} to early biological damage induced by 
ionizing radiation at the DNA scale \cite{doi:10.1142S1793962310000122}. 

A version of Geant4 with the first implementation of a 
physical process in a crystal was released with the process of phonon propagation \cite{brandt2012,2014arXiv1403.4984B}, but no 
orientational effects for charged particles were developed at that time.  The concurrent presence of many physical processes forces the use 
of an integration step greater than a ${\mu}$m to limit the computational
time.  As a result, the full solution of the equation of motion is not 
suitable.  An alternative approach would be to simulate orientational effects  
using experimental data, but such data (channeling of negative particles in
bent crystals, for example) do not currently exist.

In this paper we present a general model for the simulation of orientational
effects in straight and bent crystals for high energy charged particles.  The
model is based on the continuum potential approximation but does not rely on
the full integration of particle motion.  The model has been implemented in
Geant4, and validated against experimental data.

\section{Model}\label{chap_model}
In this section the models for channeling and volume reflection are presented.
Since they are based on the continuum potential approximation, a resume of the
Lindhard work and its range of applicability is presented here.

\subsection{Continuum approximation}\label{ch_model_gb}
The continuum approximation was developed by Lindhard to describe channeling
and its related phenomena, but can be extended to all orientational phenomena
because the same approximations hold.  Coherent effects are primary phenomena;
they govern the paths of primary particles, and not secondary ones, which are
determined by the path, as Lindhard stated \cite{Dansk.Fys.34.14}. Thus, four basic assumptions can be introduced for 
particles under orientational conditions:
\begin{itemize}
\item scattering angles may be assumed to be small.  Indeed, scattering at
      large angles implies complete loss of the original direction.
\item Because the particle moves at small angles with respect to an aligned
      pattern of atoms and because collisions with atoms in a crystal demand
      proximity, correlations between collisions occur.
\item Since the wave length of relativistic particle is small compared to the lattice 
	constant, a classical picture can be adopted.
\item The idealized case of a perfect lattice may be used as a first
      approximation.
\end{itemize}

Under these assumptions, the continuum approximation can be inferred, and the
potential of a plane of atoms $U(x)$ can be computed by taking the average of
the detailed potential along the direction of motion of the particle.

\begin{equation}
U(x) = N d_{p}\int\int_{-\infty}^{+\infty}dydzV(\bf{r})
\label{eqPotential}
\end{equation}
  
where $d_p$ is the interplanar distance, $N$ is the atomic density and
$V(\bf{r})$ is the potential of a particle-atom interaction.  By using the
screened Coulomb potential approximation for $V(\bf{r})$, the interplanar
potential becomes

\begin{equation}
U(x) = 2{\pi}N d_{p}Z_{1}Z_{2}e^{2}a_{TF}\exp\left(-\frac{x}{a_{TF}}\right)=U_{max}\exp\left(-\frac{x}{a_{TF}}\right)
\label{eqPotentialTemp}
\end{equation}

where $Z_{1}$ and $Z_{2}$ are the atomic numbers of particle and ion,
respectively, $e$ is the elementary charge, $a_{TF}$ is the Thomas-Fermi radius
and $U_{max}$ is the maximum of the potential.

The basis of the continuum approximation relies on the qualitative assumption
that many consecutive atoms contribute to the deflection of a particle 
trajectory.  Thus, for relativistic particles, the time of collision 
${\Delta{t}}\approx{\Delta}z / c$
multiplied by the momentum component parallel
to the plane of atoms, $p_{z}{\sim}p{\cos}\theta$, must be large compared to the
distance $d_z$ between atoms along the particle direction, where $p$ is particle
momentum, $p_{z}$ is the momentum component along the particle direction and 
$\theta$ is the angle between the particle direction and the crystal plane 
orientation.  Since the collision time is approximately 
${\sim}r_{min}/(v{\sin}\theta)$, where $r_{min}$ is the minimal distance of 
approach, the condition for the continuum approximation to hold is

\begin{equation}
\frac{{\Delta{z}}}{c}p{\cos}{\theta}\approx\frac{r_{min}}{\theta}{\gg}d_z
\label{eqCond}
\end{equation}

In the most restrictive form $r_{min}$ is determined by the condition that the 
transverse kinetic energy cannot exceed the transverse potential energy at 
$r_{min}$.

\begin{equation}
\frac{1}{2}p\beta\theta^2=U(r_{min})
\label{eqMinDist}
\end{equation}

Therefore, from previous equations, a condition can be derived for which the 
continuum approximation is still valid:

\begin{equation}
\frac{a_{TF}}{d_{z}\theta}\left(1-\frac{p\beta}{2U_{max}}\theta^2\right)\gg1
\label{eqCondContAppr}
\end{equation}

Two terms appear in this condition.  One refers to the Lindhard angle of 
channeling $\theta_{L}=\sqrt{2U_{max}/(p\beta)}$, which determines the maximum
angle for channeling.  The other is more interesting because it implies that
$\theta<{a_{TF}}/{d_z}\sim0.5{\AA}/1{\AA}\sim0.5$ is very large 
compared to $\theta_L$ at high energy.  Thus, the continuum approximation is
still valid for angles greater than $\theta_{L}$ as long as that particle does
not approach closer than $r_{min}$ to a nucleus.

The continuum potential approximation can be extended to regions closer than
$r_{min}$ to the atomic position by treating in more detail atomic displacement
in the structure.  In fact, since the crystal temperature is usually higher than
0 $K$ degree, atoms vibrate around their center of mass.  By averaging the 
thermal vibration amplitude over space and time, the probability density 
function for the position of atoms can be derived.  Thus, the continuum 
approximation can be extended to regions closer to the center of vibration of
atoms.  Because the averaging is due to thermal fluctuations, such an 
approximation is not valid at very low temperatures and the limits of the 
continuum approximation must be kept in mind.

\subsection{Channeling}\label{ch_model_ch}
\begin{figure}[ht]
\centering
\includegraphics[width=11cm]{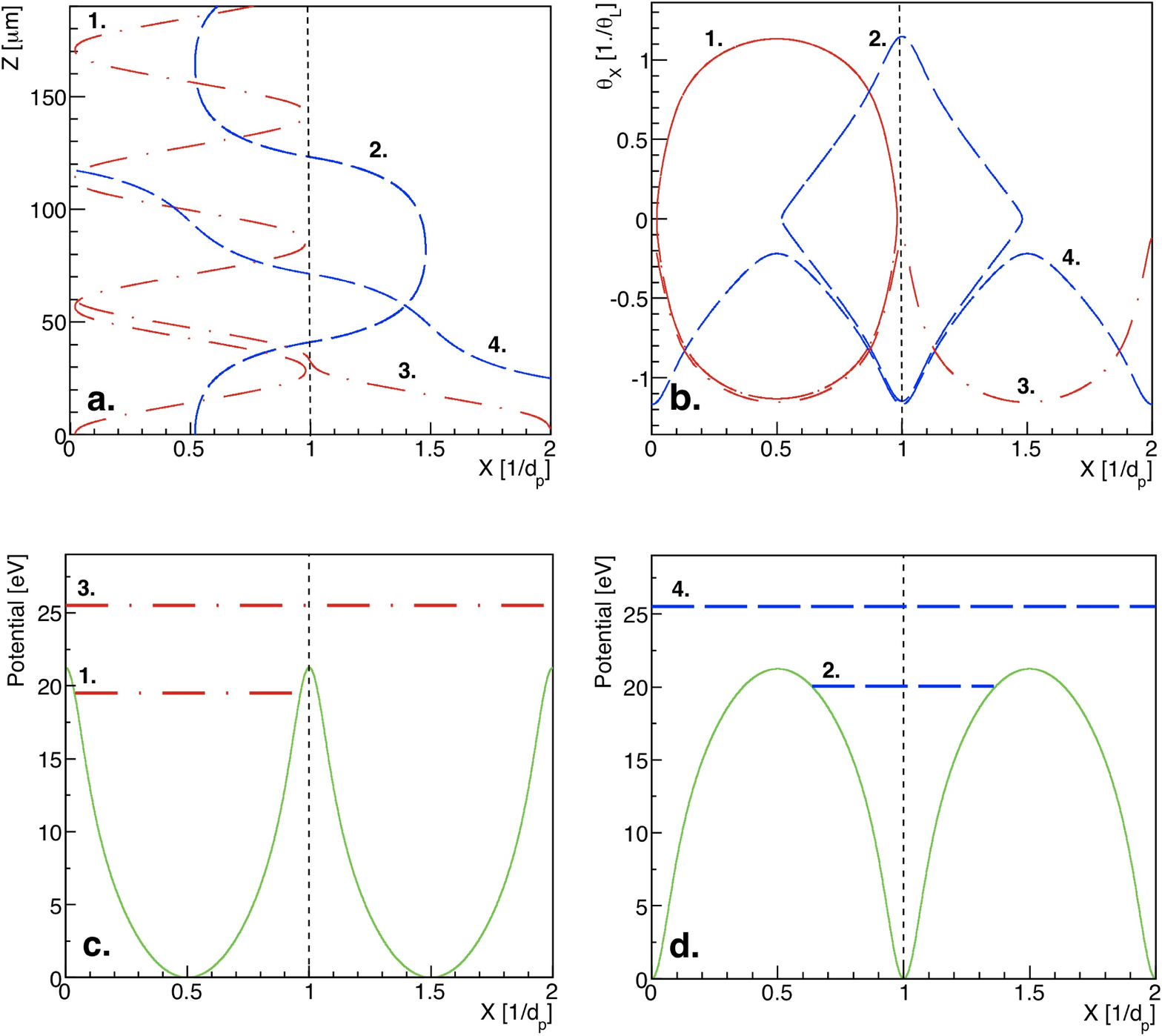}
\caption{\label{fig1} $400 GeV/c$ particles interacting with Si (110) planes 
         (dotted lines).  Curves 1 and 2 refer to channeled particles while curves
         3 and 4 refer to over-barrier particles.  Dashed (dot-dashed) lines 
         represent negative (positive) particles. (a) Trajectories as a function
         of transverse position (X) and penetration depth (Z). (b) Trajectories 
         as a function of transverse position (X) and transverse angle ($\theta_{X}$).
         Continuum planar potential (continuous line) and transverse energies for
         (c) positive and (d) negative particles.}   
\end{figure}

When a charged particle hits a crystal aligned with an atomic plane it can be
trapped by the strong electromagnetic field between two planes, thus undergoing
planar channeling.  Channeled particles follow the direction of the crystal 
plane, oscillating between or across planes if the particle charge is positive
or negative, as shown in Fig. \ref{fig1}.a.
Under channeling conditions positive particles penetrate deeper into the crystal
relative to the un-aligned orientation because the trajectory is repelled from 
the nuclei.  On the other hand, negative particles interact more frequently 
because of their attraction to zones with high densities of nuclei.

The continuum interplanar potential for main planes in crystals \cite{Biryukov}
can be approximately described by a harmonic potential well for positive 
particles, as shown in Fig. \ref{fig1}.c.  However for negative particles, 
being attracted by nuclei, the interplanar potential must be reversed and 
becomes non-harmonic with a minimum in the middle of the potential
well, as shown in Fig. \ref{fig1}.d. Because the trajectory is strongly 
affected by such a potential, positive and negative particles under channeling 
trace different shapes in phase space (see Fig. \ref{fig1}.b).

Channeling holds for particles with transverse energy $E_{x,\theta}$ lower than 
the maximum of the potential well depth $U_{0}$, i.e., $E_{x,\theta}<U_{0}$.  
Such particles follow the channeling plane or axes until they exit the crystal 
or are dechanneled.  The dechanneling mechanism behaves the same for both 
straight and bent crystals.  If all the processes which lead to dechanneling 
are disabled, a particle remains under channeling for the entire crystal length
as long as $E_{x_{in},\theta_{in}}<U_{0}$, where $x_{in}$ and $\theta_{in}$ are
the impact position and incoming angle with respect to the channeling plane.  Thus,
conservation of transverse energy allows the treatment of channeling through a 
knowledge of the initial impact position on a crystal channel $x_{in}$ and the 
angle with respect to the crystal plane ${\theta}_{in}$ :

\begin{equation}
E_{x_{in},\theta_{in}} = U(x_{in}) + \frac{1}{2}p{\beta}{\theta_{in}}^{2} .
\label{eqEnergy}
\end{equation}

Solving the equation of motion requires point-by-point knowledge of the 
transverse position and transverse momentum of an oscillating particle.  
However, by choosing a crystal which extends along the beam for more than one 
oscillation period of a channeled particle, the energy level occupied by a 
particle in the electrostatic potential well generated between atomic planes 
or axes is the only physical quantity to link initial to final parameters in a 
real-case study.  By imposing a continuous and uniform distribution in position
$x_{out}$ for a channeled particle of energy $E_{t}$, an outgoing angle 
$\theta_{out}$ is generated by evaluating

\begin{equation}
\theta_{out} = \sqrt{2p{\beta}(E_{t} - U(x_{out}))}
\end{equation}

Therefore, information regarding $x_{in}$ and ${\theta}_{in}$ can be condensed 
into a single variable $E_{t}$, which determines the occupied energy level of a 
channeled particle and allows the computation of the outgoing distribution of 
channeled particles by means of $E_{t}$ and the continuum potential.

For bent crystals the model is still valid.  The sole difference relies on the 
modified potential in the non-inertial reference frame orthogonal to the crystal
plane or axis.  In fact, the centrifugal force acting on the particle in this frame 
pulls down the potential barrier resulting in a shallower potential well.
Thus, the condition for channeling holds with a modified maximum potential and 
transverse energy related to the non-inertial reference system.

The presence of torsion in a crystal spoils channeling efficiency in bent 
crystals \cite{ipac10tors}.  Indeed, the orientation of the channeling angle 
with respect to the beam direction changes with the impact position on the 
crystal surface.  Since a beam has a finite size, two particles with the same 
direction and the same impact position on the potential well but different 
impact positions on the crystal surface have different transverse energies.  
This effect is introduced in the simulation by changing the plane direction with
respect to the impact position on the crystal surface.

Another important parameter for channeling in bent crystals is the miscut
\cite{Tikhomirov:1384907}, which is the angle between the lateral surface of a 
crystal and the atomic planes.  Only the trajectories of particles channeled 
near a crystal edge are affected by the presence of the miscut, because it 
modifies the total length of the bent plane of channeling.  This effect is 
introduced by defining the plane orientation independently of the crystal 
volume.

\subsection{Dechanneling and volume capture}\label{ch_model_dech}

Particles which no longer satisfy the channeling condition have suffered 
dechanneling.  Unchanneled particles which enter the channeling state undergo
volume capture.  Dechanneling and volume capture take place when particles 
interact incoherently with nuclei or electrons.  Indeed, a channeled particle 
can acquire enough transverse energy to leave the channeling state by exceeding
the maximum of the potential well, or an unchanneled particle can lose energy
and decrease its transverse energy by passing under the maximum of the potential
well.

All physical phenomena occurring for a channeled particle are strongly affected
by the occupied energy levels.  As shown in Fig. \ref{fig3}, the average 
density of material seen by a particle traversing a crystal aligned with its 
planes is strongly affected by the transverse energy of the particle.  Thus, the
probability of interaction with nuclei and electrons has to be weighted as a 
function of the transverse energy.  Kitagawa and Ohtsuki \cite{PhysRevB.8.3117}
demonstrated that there is a linear dependence between the incoherent 
interaction rate and the material density.  Therefore, the modified cross 
section ${\sigma}(E_{t})$ of each phenomenon is
\begin{equation}
{\sigma}(E_{t}) = {\sigma}_{am}\frac{n(E_{t})}{n_{am}}
\end{equation}
where ${\sigma}_{am}$ is the cross section in amorphous material, ${n_{am}}$ is 
the density of the amorphous material and ${n(E_{t})}$ is the modified density.

Since each incoherent phenomenon produces a variation of transverse energy
${\Delta}{E_{t}}$, the maximum distance traveled by a channeled particle is
limited by the probability of
passing over or under 
the potential barrier
during each step.  As an example, the modified rms ${\sigma}_{is}(E_{t})$ for 
incoherent scattering on nuclei depends approximately on the square root of the
traversed material density \cite{RevModPhys.13.240}. Thus, the step can be limited by the condition
\begin{equation}
|U_{0}-{E_{t}}|=|\Delta{E}|={\Delta}\left({p\beta}{\theta}^2\right)\approx{p\beta}\frac{n(E_{t})}{n_{am}}\Delta\left({\sigma}_{is,am}^2\right)
\end{equation}
where ${\sigma}_{is,am} \sim \frac{13.6 MeV}{p\beta}\sqrt{\frac{z}{X_0}}$ is the 
rms of the incoherent scattering in the amorphous material,
${\beta}$ is the particle velocity in units of the speed of light and $X_0$ is 
the radiation length of the material.  Thus, the step ${\Delta}z$ is

\begin{equation}
{\Delta}z  \sim X_{0}\frac{p\beta}{E_s^{2}}\frac{n_{am}}{n(E_{t})} |{\Delta}E|
\end{equation}

Such an approach can be applied to all the concurrent incoherent processes to 
determine the maximum step size by comparing the contributions.  The model is 
still valid for positive and negative particles since no restriction has been 
applied.

The same response for the interaction probability is not obtained by averaging 
the density over an oscillation period.  Therefore, this model must be adapted
for crystals with lengths along the beam greater than one oscillation period.
On the contrary, by integrating the particle trajectory it is possible to 
determine the interaction probability for each step depending on the position 
in the channel.  Thus,  the peculiar characteristic of channeling in the first 
layers of a crystal can not be described
by the averaging used in the model developed in this paper.

\subsection{Volume reflection}\label{ch_model_vr}
When charged particles cross a bent crystal tangent to its planes they are
"reflected" in the direction opposite to the bending curvature.  This is called
volume reflection.  In fact, the particle is deflected by the continuous
potential barrier of one plane, but immediately leaves the channel because the 
barrier of the opposite plane is lowered due to bending, and thus the particle
cannot be trapped under channeling.  Therefore, the condition for volume 
reflection holds when the projection of the particle momentum on the direction
of a plane changes sign.  Volume reflection and related phenomena limit the 
maximum allowed step length.  Indeed, particles can be captured into a 
channeling state if they lose enough transverse energy to fulfill the channeling
condition $E_{x}<U_{0}$.  Thus, the step length must be comparable to the 
oscillation period near the turning point.  The distance of a particle to the 
tangency point in a bent crystal must be evaluated at each step to set the step
size at the proximity of the interesting region.

\begin{figure}[ht]
\centering
\includegraphics[width=11cm]{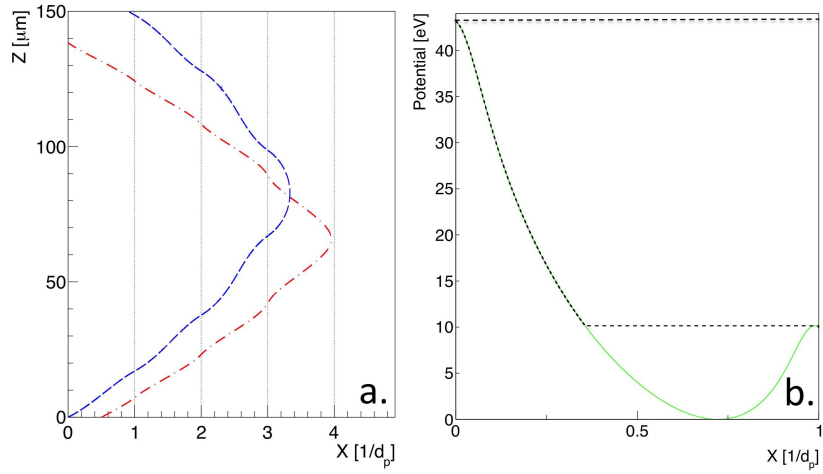}
\caption{\label{fig2} (a) Simulation of volume reflection for positive 
         (dot-dashed line) and negative (dashed line) particles with the same
         initial transverse energy in the non-inertial reference frame orthogonal
         to the crystal plane.  Dotted lines are crystal planes.  The higher 
         momentum of positive particles near the turning point results in a
         larger deflection angle for volume reflection. (b) Continuum planar 
         potential in the non-inertial reference frame orthogonal to the Si (110) 
         plane for $p\beta/R=17.3eV/{\AA}$.  Dotted lines delimit region of
         volume reflection.}
\end{figure}

For a slightly bent crystal, the mean deflection angle for volume reflection is 
approximately $\sim1.4\theta_{L}$ for positive particles
\cite{PhysRevLett.101.234801} and $\sim0.8\theta_{L}$ for negative particles
\cite{Taratin1987512}, where $\theta_{L}=(2U_{0}p\beta)^{1/2}$ is the Lindhard
angle \cite{Dansk.Fys.34.14}.  In fact, a positive particle spends more time in
a zone within which the particle has higher transverse velocity, while the 
opposite is true for a negative particle, as shown in Fig. \ref{fig2}.a.  By 
decreasing the radius of curvature, the mean deflection angle decreases.
Indeed, the deflection angle of volume reflection depends on the transverse 
energy at the turning point.  The more the crystal is bent, the larger is the
angular spread \cite{PhysRevLett.101.234801}.  In Fig. \ref{fig2}.b the 
potential in the non-inertial reference frame orthogonal to the crystal plane
is shown.  The maximum energy difference delimits the reflection region.  The
potential shapes the trajectories of particles with transverse energies a bit
above the maximum of the potential.  These are the so-called over-barrier 
particles.  By approximating the potential in the reflection region with a 
linear function, the volume reflection angle becomes proportional to the 
position of the turning point.  Thus, the deflection angle can be generated by
adopting a continuous and uniform distribution proportional to the reflection
region of the potential barrier (see Fig. \ref{fig2}.b).

\subsection{Average density}\label{ch_model_md}
\begin{figure}[ht]
\centering
\includegraphics[width=11cm]{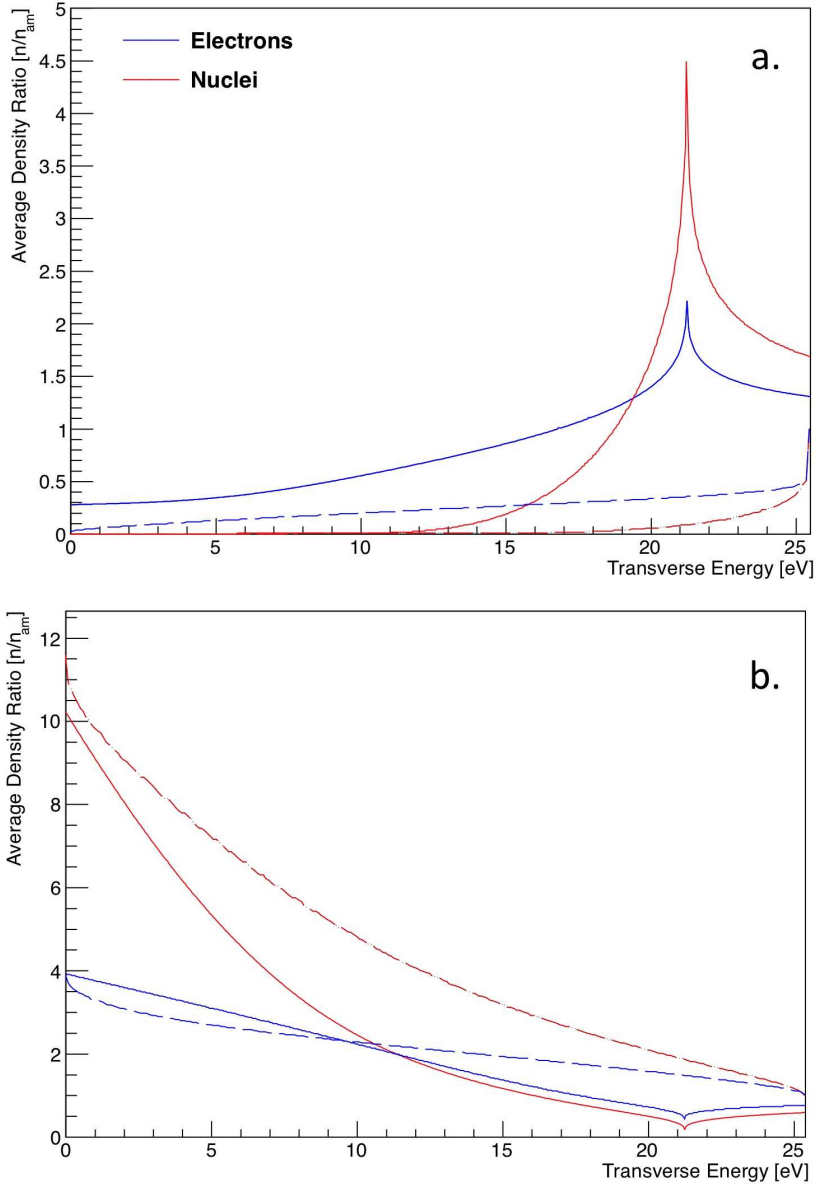}
\caption{\label{fig3} (a) Ratio of average density of nuclei and electrons
         to density of an amorphous material as a function of the transverse
         energy of a positive channeled particle. Continuous lines represent 
         the DYNECHARM++ calculation while dashed lines represent the fast model.
         Over-barrier particles experience greater density as a result of 
         different motion in the cryatal lattice. (b) Same as (a) but for a 
         negative channeled particle.  Note the difference in vertical scales.} 
\end{figure}

The average density seen by a particle undergoing orientational effects is a 
very important parameter for the model proposed in this paper.  As shown in 
Fig. \ref{fig3} the average density is strongly affected by the transverse 
energy for channeled and over-barrier particles.  The computation is made with 
the DYNECHARM++ code \cite{Bagli2013124} in which all incoherent processes are 
disabled.  The DYNECHARM++ code is based on the full solution of the equation 
of motion in the continuum potential and allows the computation of electric 
characteristics of the crystal through the ECHARM calculation method 
\cite{PhysRevE.81.026708,ipac10ECHARM}.  Therefore, the density as a function 
of transverse energy for complex atomic structures and for many planes and axes
can be computed.

The calculation of average density by DYNECHARM++ is very accurate,
but can be very slow.
Thus, a fast version has been developed to compute
the average density.  By integrating the density of all the possible states in 
which a particle with transverse energy $E_t$ can exist, the approximate average
density $\mathrm{\overline{\rho}}(E_{t})$ can be computed.
 
\begin{equation}
\mathrm{\overline{\rho}}(E_{t})=\int_{U(x)<E_{t}}{\rho}(x)dx
\end{equation}

Since no approximation was imposed on $U(x)$ and ${\rho}(x)$, this approximation
is still valid for any potential and average density function.  Experiments with
orientational effects rely mostly on the use of crystal planes with low Miller
indexes.  For these the averaged nuclei density $\mathrm{\overline{\rho}}(x)$ is
analytically derived starting with the density of nuclei function averaged over
thermal and space fluctuations

\begin{equation}
\rho(x)=\frac{1}{u_T\sqrt{2\pi}}e^{-\frac{x^2}{2u_T^2}}
\end{equation}

where $u_t$ is the thermal vibration amplitude.  Thus, the average density is
 
\begin{equation}
\mathrm{\overline{\rho}}(E_{t})={\tt{erf}}(x | U(x)=E_{t}) + 1
\end{equation}

where $E_{t}\leq U_{0}$ and $U(x)=U_{0}$ at $x=d_{p}/2$.  Fast calculation 
models do not take into consideration the time spent in a particular region by 
a particle.  The model can be applied to compute density for both positive and 
negative particles.  In Fig. \ref{fig3} the average density ratio of 
electrons and nuclei is shown for both the fast model and DYNECHARM++.

\section{Geant4 implementation}\label{ch_geant4}
The Geant4 toolkit allows new physical processes to be added to the standard 
ones it already provides. Thus, a process  can be added to 
already developed simulations with minor modification of the code.
As a consequence, the influence of the new process on existing experimental apparatus can be studied.
As an example, with the addition of the channeling process, the influence of channeling on the production
of secondary particles in a crystal collimation scheme as well
 as in a crystal extraction scheme can be simulated.

A new process must provide its mean interaction 
length and how particle properties are affected by the interaction. 
Indeed, at each step, the toolkit computes for all the processes their mean interaction length and
 the shortest one limits the maximum step 
the particle can traverse in a geometrical volume.
If the process occurs, the particle parameters are modified by  the process.
Then, the particle moves to the new position and the routine takes place for a new step

The model proposed in this paper has been implemented by a process describing the 
orientational process, and wrappers that modify the material density
in existing processes. In addition, the capability of calculating the crystal electrical
characteristics have been inserted to allow the simulation of orientational 
processes with no need for external software.

\subsection{Channeling process}
The class used for the implementation of orientational
processes is called $\tt{ProcessChanneling}$. It inherits from the virtual class 
which defines the behavior of discrete physical phenomena ($\tt{G4VDiscreteProcess}$ class).
Because the particles may undergo channeling only in a crystal,
 the channeling process is valid only in a volume with a crystal lattice.

When a particle crosses the boundary between two geometrical volumes, one with
and one without a crystal lattice, the channeling process limits the step of the
particle and checks if the particle is subject to orientational effects.  A 
uniformly distributed random number is generated to determine the impact
position of the particle on the crystal channel $x_{in}$ and, consequently, to 
compute the initial potential energy $U_{0}$.  The particle momentum is 
projected on the channeling plane to evaluate the transverse momentum.  The 
initial transverse energy $E_{x_{in},\theta_{in}}$ is computed through 
Eq. \ref{eqEnergy}.  Thus, $E_{x_{in},\theta_{in}}$ is
used
to find the 
modified density $\overline{\rho}(E_{t})$.  If the particle satisfies the 
channeling condition, $E_{x_{in},\theta_{in}}<U_{0}$, the channeling process 
proposes to the Geant4 core an alignment of the particle momentum with the 
direction of the channeling plane.  The condition for channeling is recomputed 
until the particle exits the volume with the crystal lattice.

Volume reflection occurs only for bent crystals under the condition defined in 
section \ref{ch_model_vr}.  Under volume reflection, the particle momentum 
vector is rotated by the volume reflection angle around the axis orthogonal to
the channeling plane.

\subsection{Crystal}
The class for the description of a crystal structure ($\tt{XVPhysicalLattice}$ class) was introduced into Geant4.
 In order to define a geometrical volume as a crystal,
the class has to be attached to a physical volume. 

This class collects the crystal data, such as unit cell ($\tt{XUnitCell}$ class) 
and bases ($\tt{XLogicalBase}$ class). The base contains the kind and disposition of the atoms. 
The unit cell groups the unit cell information, i.e., the sizes and
the angles of the cell, and holds a vector of pointers to as many 
bases as needed.  The information stored in a 
unit cell may be used to compute electrical characteristics under 
the continuum approximation of the channeling processes.

\subsection{Wrappers}
At each step in a crystal, the particle momentum can be modified by any of the
Geant4 processes.  Such modifications vary the transverse energy of a particle
and may cause dechanneling, that is, the overcoming of the potential well 
maximum.  As stated in section \ref{ch_model_dech}, the average densities of 
nuclei and electrons change as a function of the transverse energy of the 
particle.  Thus, these densities should be recomputed at each step and used to 
modify the cross section of the physics processes which depend on the traversed
quantity of matter (see section \ref{ch_model_md} ).

In order to modify the
cross section
of existing processes and to preserve code
re-usability for future releases of Geant4, wrapper classes for the 
discrete and continuous processes were developed.  For both these 
classes, the interaction length of discrete processes is resized proportionally 
to the modified material density.  For the energy loss of the 
continuous processes, the traversed length is resized
in proportion to 
the 
modified average density.  For each wrapped process a wrapper object must be 
instantiated.  The wrappers need only the average density to recompute the
process cross section.  Thus, in principle, it may work independently of the 
channeling process.

\section{Examples of calculation}\label{ch_valid}
Model validation has been completed by comparison with published experimental 
data.  Experiments studying the efficiency of channeling vs. incoming angle
\cite{Scandale2009129}, the rate of inelastic nuclear interaction under channeling
\cite{Scandale20102655}, and the channeling efficiency dependence on radius of 
curvature for bent crystals \cite{doi.org/10.1140/epjc/s10052-014-2740-7}, were
simulated for positive particles.  For negative particles, simulations of the 
dechanneling length for
high energy
pions \cite{Scandale201370} was performed.
Comparing simulations to experiments allowed both the precision of the model and
the quality of the Geant4 implementation to be checked.

A bent crystal was modeled as a small fraction of a toroid with a bending radius
on the order of a meter and a length on the order of a mm along the beam 
direction, matching the dimensions used in the experiment.  Though torsion can 
be simulated, it was set to zero for all the current simulations.  This has
no effect on the agreement of simulation with data, even though the  
experimental data have been corrected for torsion.  In addition, the miscut 
value has no influence because only particles impinging far from crystal edges 
have been used in the analyses.

As in the experimental setups, three silicon detectors were inserted into the
simulation along the beam direction to track the particle.  For measurement of
the rate of inelastic nuclear interaction, two scintillators were added to 
reproduce the experimental setup of Ref. \cite{Scandale20102655}.  To speed up
simulation, volumes other than crystal and detectors have been filled with 
galactic vacuum ($\tt{G4\_Galactic}$ material).

\subsection{Positive particles}
\begin{figure}[ht]
\centering
\includegraphics[width=11cm]{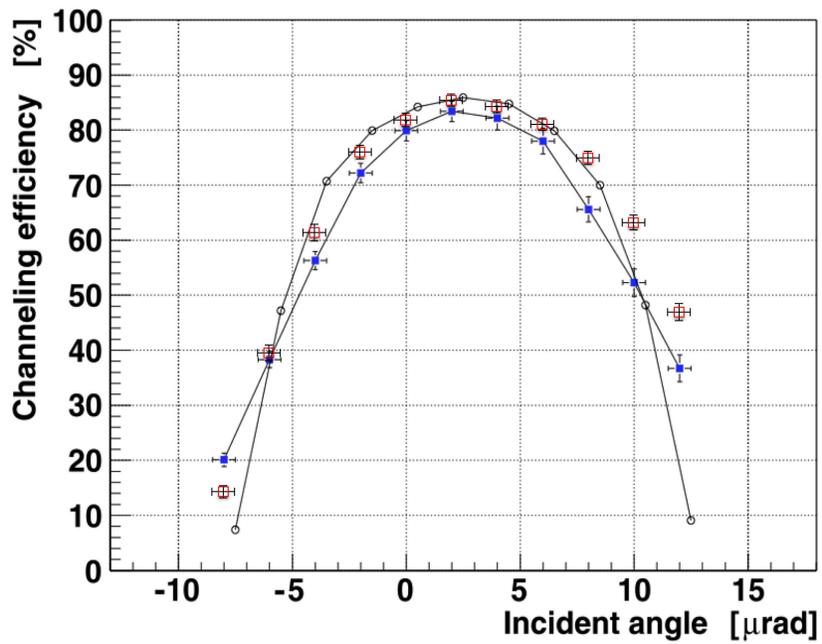}
\caption{\label{fig4} Deflection efficiency for a narrow 
         beam as a function of the incoming beam direction with respect to plane direction
         of a (110) Si bent crystal.  Empty squares are the results of a Geant4 
         simulation with the model described in the paper, filled squares are
         experimental measurements and circles are simulations with complete
         integration of the particle trajectory.  The figure is partially a
         reproduction of Fig. 3 of Ref. \cite{Scandale2009129}.  Filled blue
         squares are experimental data, open red squares are Geant4 simulation,
         black open circles are Monte Carlo simulations with full integration
         of the trajectories.}
\end{figure}

In Fig. \ref{fig4}, the channeling efficiency as a function of incoming 
angle is superimposed on experimental results (Fig. 3 of 
Ref.  \cite{Scandale2009129}) and a Monte Carlo simulation with complete 
integration of the trajectories.  The maximum efficiency for channeling in 
Geant4 is in good agreement with experimental data as well as efficiency in the
tails.  However, fair agreement is obtained in the region between maximum 
efficiency and tail, with $\sim5\%$ deviation in efficiency.  In this region 
the model lacks accuracy because the trajectories were not completely 
integrated.  Thus, such behavior is to be ascribed to the shape of the 
interplanar potential used in simulation for both the models.

\begin{figure}[ht]
\centering
\includegraphics[width=11cm]{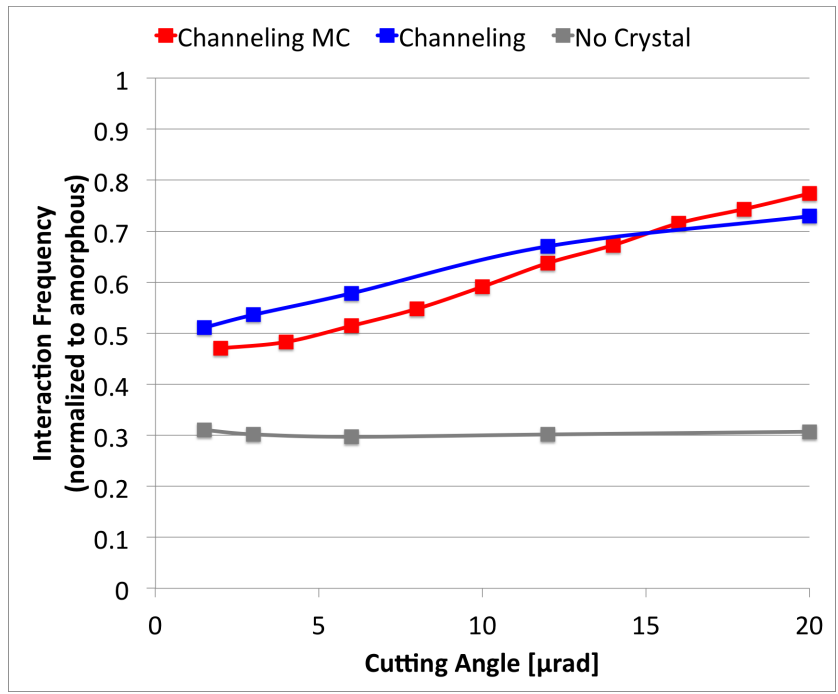}
\caption{\label{fig5} Dependence of the inelastic nuclear interaction rate
          of protons on
          the beam angular spread of a 400 GeV/c incident proton beam channeling (blue line). Monte Carlo simulation with Geant4 are superimposed (red line). Gray line shows the background measurement with no crystal along the beam. Experimental data have been taken from Fig.5 of Ref. \cite{Scandale20102655}.}
\end{figure}

Fig. \ref{fig5} should be compared with Fig. 5 of Ref. 
\cite{Scandale20102655}.  The rate of secondary particles as a function
of the beam angular spread is shown normalized to an amorphous condition.  The standard Geant4 release without the channeling 
extension has been used for simulations with amorphous Si and with no crystal.
Simulations are in agreement with experimental data.  The channeling extension
allows the correct modification of the cross sections of incoherent phenomena,
reducing the rate with respect to amorphous materials.  Discrepancies are 
observed for small angles and the slope of the two curves are different.

Table \ref{tab1} presents the deflection efficiency for channeling vs. radius of
curvature.  Experimental data and DYNECHARM++ simulations are taken from 
Ref. \cite{doi.org/10.1140/epjc/s10052-014-2740-7}.  As the critical radius 
$R=R_c$ is approached, the discrepancy between experimental data and simulation
increases.  Such behavior is also recorded for DYNECHARM++ simulations.  As 
stated in Ref. \cite{doi.org/10.1140/epjc/s10052-014-2740-7}, the discrepancy 
must be ascribed to the lack of knowledge of the exact density distribution 
between atomic planes.  An important feature of Geant4 is its capability to 
evaluate the number of particles which suffer nuclear interaction or are 
scattered at large angles.  In Table \ref{tab1} the fraction of "lost" particles
, i.e., which do not hit the last detector, is reported. 

\begin{table}[!ht]
\caption{\label{tab1} Measured channeling efficiency ($\%$) (Exp.), and simulated
 efficiency calculated with Geant4 (G4) and with DYNECHARM++ (D++) methods, and 
 the fraction of particles which do not hit the last detector for the Geant4
 simulation (G4 (lost)).}
\begin{center}
\begin{tabular}{ c | c c c c}
$R/R_c$ & Exp. & G4 & G4 (lost) & D++\\ \hline
40.6  & 81 & 84 & 0.8 & 81.2\\  
26.3 & 80 & 81 & 0.8 & 79.7\\  
9.7 & 71 & 75 & 0.8 & 72.3\\  
5.1 & 57 & 61 & 0.9 & 56.8\\  
3.3 & 34 & 44 & 1.0 & 39.9
\end{tabular}
\end{center}
\end{table}%

\subsection{Negative particles}
\begin{figure}[ht]
\centering
\includegraphics[width=11cm]{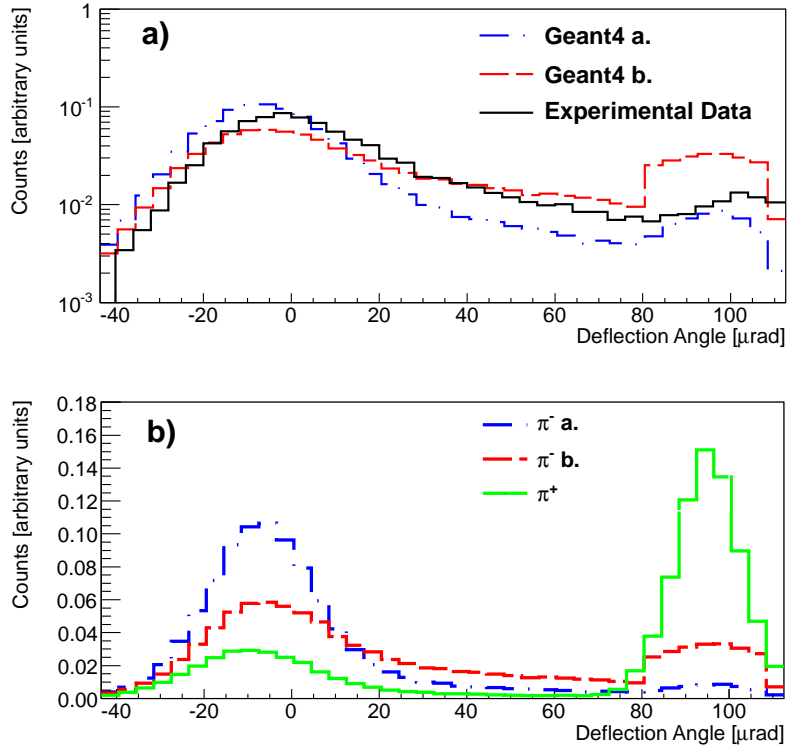}
\caption{\label{fig6} a) Geant4 simulation of the distribution of deflection
         angle of $150$ GeV/c $\pi^-$ passed through a $1.91$ mm long silicon
         crystal bent at $R$ = $19.2$ m along (110) planes.  Only particles
         hitting the crystal within an angle of $5$ $\mu$rad are selected.
         The average density experienced by a channeled particle has been computed
         by DYNECHARM++ (Geant4 a.) and an algorithm implemented in Geant4
         (Geant4 b.).  Experimental data was published in Ref. \cite{Scandale201370}.
         The dechanneling length was evaluated with the method proposed in the same 
         reference. b) Geant4 simulation for the same crystal with $150$ GeV/c 
         $\pi^-$ (a. and b.) and $150$ GeV/c $\pi^+$.}
\end{figure}

In Ref. \cite{Scandale201370} the interaction of $150$ GeV/c negative pions with
a bent Si crystal has been studied in order to measure the dechanneling length 
for negative particles.  A dechanneling length of $1.54\pm0.05$ mm was obtained
by Geant4 simulation with the density computed by DYNECHARM++ code, compared to
$0.71\pm0.05$ mm with the density computed by the new Geant4 model (see 
Fig. \ref{fig6}.a).  The dechanneling rate is increased due to the stronger
incoherent scattering with nuclei and electrons.  Thus, the model for negative 
particles is very sensitive to the interaction rate in one oscillation period, 
since a big discrepancy between the two simulations exists.  Indeed, the 
discrepancy of the dechanneling lengths becomes large for the channeling 
efficiency, which goes from $26.8\pm0.5\%$ to $6.2\pm0.5\%$.  As a 
consequence, by computing accurately the average density experienced by a 
particle the model is able to output the measured dechanneling length.

The same configuration was used to simulate channeling of $150$ GeV/c $\pi^+$.
The comparison between positive and negative pions is shown in 
Fig. \ref{fig6}.b.  The deflection efficiency for $\pi^+$ is $\sim70\%$, which
is greater than than for $\pi^-$.  This result demonstrates that the channeling
model developed for Geant4 allows positive and negative particles to be managed 
differently thanks to the wrapper classes.

\subsection{Computation time}\label{ch_comp}
The Geant4 code has been compared to the DYNECHARM++ code in order to evaluate 
advantages of the approach proposed in this paper in terms of computation.  The
same initial conditions have been used as in Ref. \cite{Bagli2013124} : a $400$
GeV/c proton beam interacting with a $1.94$ mm thick (1 1 0) Si bent crystal 
with a $38$ m radius of curvature.  The Geant4 single-threaded version 10.00b 
has been adopted and only a discrete single scattering model \cite{Mendenhall2005420} has been added to
its list of physics processes.  The computer was the same as that used for 
DYNECHARM++ test, i.e. a personal computer with 8 GB of RAM and an Intel(R) 
Core(TM) i7-2600K CPU running at 3.40GHz.  Computation time was approximately 
14 ms per particle in Geant4 vs. 38 ms per particle in DYNECHARM++, in spite of
the greater complexity of the Geant4 code.  This result is explained by 
considering the number of steps required by the two models adopted for the 
simulation.  Full integration of trajectories requires step sizes much smaller
than the oscillation period in the potential well.  On the contrary, the 
Geant4-based model allows the use of a step size comparable to the oscillation 
period.

\section*{Conclusions}
The exploitation of orientational processes in crystals to manipulate particle
trajectories is currently a topic of intense interest in physical research, with
possible applications for the LHC for beam collimation \cite{Scandale:1357606} 
and extraction \cite{Bellazzini:315432,Rakotozafindrabe:1492806,Lansberg:1503414}.  A physical 
model suitable for the Monte Carlo simulation of such processes has been 
developed.  This model relies on the continuum potential approximation.  The 
model makes use of the transverse energy in the non-inertial reference frame 
orthogonal to the channeling plane in order to discriminate between channeled 
and unchanneled 
particles.  The average density experienced by a channeled 
particle is evaluated in order to compute the modification of the cross section
for hadronic and electromagnetic processes.  The model represents an extension 
of the Geant4 toolkit.  The code has been validated against data collected by 
experiments at CERN.  It demonstrates that Geant4 is able to compute 
the deflection efficiency for channeling and the variation of the rate of 
inelastic interactions under channeling.

\section*{Acknowledgments}
We acknowledge partial support by the INFN under the ICERAD project and by the Sovvenzione Globale Spinner 2013 grant 188/12 with the ICERAD-GEANT4 project.
\bibliographystyle{spphys}
\bibliography{biblio}

\newpage

\section*{Figure}

\end{document}